\documentclass[manuscript]{aastex}

\usepackage{lscape}

\usepackage[usenames,dvips]{color}
\usepackage{graphicx}

\shorttitle{Population study for $\gamma$-ray
pulsars}
\shortauthors{Takata J. et al. }

\begin{document}


\title{Population study for $\gamma$-ray pulsars with the outer gap model}


\author{J. Takata\altaffilmark{1}, Y.Wang\altaffilmark{2} and
K.S. Cheng\altaffilmark{3}}
\affil{Department of Physics, University of Hong Kong,
Pokfulam Road, Hong Kong}

\altaffiltext{1}{takata@hku.hk}
\altaffiltext{2}{yuwang@hku.hk}
\altaffiltext{3}{hrspksc@hkucc.hku.hk}


\begin{abstract}
 Inspired by  increase of  population of $\gamma$-ray emitting pulsars 
 by the $Fermi$ telescope, we perform a population study  for  $\gamma$-ray 
emitting canonical pulsars. We use a Monte-Carlo technique to simulate 
the Galactic
 population of neutron stars and the radio pulsars. For each simulated 
neutron star, we consider the $\gamma$-ray emission from the outer gap 
accelerator in the magnetosphere. In our outer gap model, 
we apply the gap closure mechanism proposed by Takata et al., in which both 
 photon-photon pair-creation and magnetic pair-creation processes are considered. 
 Simulating the sensitivities of  previous  major  radio surveys, 
our simulation predicts that there are $\sim 18-23$ radio loud   
 and $\sim 26-34$ $\gamma$-ray-selected 
  $\gamma$-ray pulsars, which can be detected 
with a $\gamma$-ray flux $F_{\gamma}\ge 10^{-10}~\mathrm{erg/cm^2 s}$.
Applying the sensitivity of the six-month observation of the $Fermi$ 
telescope, 40-61 radio-selected and 36-75
 $\gamma$-ray selected pulsars are  detected within our simulation. 
 We show that the  distributions of  various pulsar parameters for the
 simulated $\gamma$-ray pulsars can be consistent with  
the observed distribution of the $\gamma$-ray pulsars detected by 
the $Fermi$ telescope.  We also predict that $\sim 64$ radio-loud 
and $\sim 340$ $\gamma$-ray-selected pulsars  irradiate the Earth 
with a flux  $F_{\gamma}\ge 10^{-11}~\mathrm{erg/cm^2 s}$,
 and  most of   those 
 $\gamma$-ray pulsars are distributing with a distance more than 1~kpc 
and a flux $F_{\gamma}\sim  10^{-11}~\mathrm{erg/cm^2 s}$.  The ration between 
the radio-selected and $\gamma$-ray-selected pulsars depend on the sensitivity 
of the radio surveys. 
 We also discuss the Galactic distribution of the unidentified $Fermi$ 
sources and the canonical $\gamma$-ray pulsars.  
\end{abstract}





\section{Introduction}
The $Fermi$ $\gamma$-ray telescope has drastically increased  population 
of $\gamma$-ray emitting pulsars after its launch in June, 2008. Abdo et al. 
(2010a) reported the first $Fermi$ pulsar catalog, including 21 radio selected 
pulsars, 16 $\gamma$-ray-selected pulsars,
the Geminga and 8 millisecond pulsars (see also  Abdo et al. 2009a,b), 
while the pulsed radio emissions are reported from several $\gamma$-ray-selected
 pulsars (Camilo et. al. 2009).  Saz~Parkinson et al. (2010) have also 
 reported new eight gamma-ray pulsars discovered in blind searches of $Fermi$-
LAT data.  These $Fermi$ observations for the  $\gamma$-ray emitting pulsars
 have started to break the bottleneck of study of the high-energy radiation 
from the pulsar magnetosphere.  

The particle acceleration and resultant high-energy emissions 
from the pulsar magnetosphere have been studied  
with  polar cap  model
(Ruderman \& Sutherland  1975; Daugherty \& Harding 1982, 1996),
 slot gap model (Arons 1983; Muslimov \& Harding  2003; Harding
et al. 2008) and outer gap model (Cheng, Ho \& Ruderman 1986a,b; Hirotani 2008; Takata, Wang \& Cheng 2010; Wang, Takata \& Cheng 2010). All emission 
models assume that the electrons and/or positrons are accelerated 
by the electric 
field along the magnetic field lines and that the accelerated particles emit 
the  $\gamma$-rays either the curvature radiation 
or the inverse-Compton processes. 
The polar cap model assumes the emission region close to the stellar surface 
and above the polar cap region, while the outer gap and slot gap models
 assume an acceleration region extending to the outer magnetosphere.

Properties of the pulse profiles and spectra of the $\gamma$-ray 
emission measured by the $Fermi$ 
can be used to discriminate between the different models.   
For example, most of the  observed pulse profiles of the $\gamma$-ray pulsars 
detected by $Fermi$ show the double peak structure with a phase-separation of 
$~\sim 0.4-0.5$, indicating  a high-altitude emission 
from the outer gap or slot 
gap region is favored more than the low altitude emission from the polar cap accelerator (Romani \& Watter 2010; Venter, Harding 
\&  Guillemot 2009). 
Furthermore,  in the
first $Fermi$ pulsar catalog (Abdo et al. 2010a), the spectral fits have
been done assuming exponential cut-off function. It is found that
the observed  spectrum for the Vela pulsar (Abdo et al. 2009c, 2010b) 
is well fitted by a power
 law plus hyper-exponential
cutoff spectral model of the form $dN/dE\propto E^{-a} \exp[-(E/E_c)^b]$
with $b\le 1$ and $E_c\sim 1.5$~GeV.
 This implies that an  emission process in
 the outer magnetosphere is  favored more than that near the stellar
 surface,  which predicts a super exponential
cut-off ($b>1$) due to the magnetic pair-creation prosses of
 the $\gamma$-rays.  In addition, the detection of the radiation above
25~GeV bands associated with the Crab pulsar by the MAGIC telescope
has also implied the
emission process in the outer magnetosphere (Aliu et al. 2008).

The increase of population of the $\gamma$-ray emitting pulsars also allow us
 to perform a more detail study for the high-energy emissions from the 
pulsars. Wang et al. (2010) 
fit the observed phase-averaged spectra using a  two-layer 
 outer gap model, and then investigated  the relation between the $\gamma$-ray 
luminosity ($L_{\gamma}$) and the spin down power ($L_{sd}$) for
 39 $\gamma$-ray emitting pulsars including 8 millisecond pulsars. 
They suggested  that the relation can be expressed 
as $L_{\gamma}\propto L_{sd}^{\beta}$ with $\beta\sim 0$ for 
$L_{sd}\gtrsim 10^{36}$~erg/s, while $\beta\sim 0.5$
 for $L_{sd}\lesssim 10^{36}$~erg/s. They further argued that the relation 
between the $\gamma$-ray luminosity and the spin-down power implies that 
the outer gap is closed by the photon-photon pair-creation process for
   $L_{sd}\gtrsim 10^{36}$~erg/s (Zhang \& Cheng 1997), while by 
 magnetic pair-creation process for  $L_{sd}\lesssim 10^{36}$~erg/s 
(Takata et al. 2010). 

For the outer gap model, the population studies were done to compare the 
model prediction with the results of $EGRET$ instrument (Cheng \& Zhang 1998; 
Zhang et al. 2004; Cheng et al. 2004). Applying the photon-photon 
pair-creation model for gap closure process,  they found that the 
distributions of the various pulsar parameters 
(e.g. rotation period, spin down age) were consistent with the observations. 
Gonthier et al. (2002) and Story, Gonthier and Harding (2007) studied the 
population of $\gamma$-ray pulsars with the polar cap and 
slog gap accelerator models to predict the results using 
 the $Fermi$ observations.
 Furthermore, Grenier \& Harding (2010) has studied the distribution of the 
$\gamma$-ray pulsars predicted by 
 the outer gap model of  Zhang et al. (2004) and for the slot gap model 
 of  Muslimov \& Harding (2003; 2004) to compare with the $Fermi$ results.  
They have argued that both
 outer gap and slot gap model  may not explain all features of the $Fermi$
pulsars; for example,  both models predict too few young pulsars
compared with the $Fermi$ pulsars.

 In this paper,  we will further develop the population study
 for the $\gamma$-ray emitting pulsars to compare with the results
 of six-month $Fermi$ observations. 
In particular,  we will consider  the $\gamma$-ray emissions of the canonical 
$\gamma$-ray pulsars with the  outer gap model. We will apply the 
 gap closure model proposed by Takata et al. (2010),
 in which the photon-photon pair-creation process 
and the magnetic pair-creation process are taken into account. As usual, 
we will perform a Monte-Carlo simulation for the Galactic population 
of the neutron star (section~\ref{simulation}).   In section~\ref{gemission}, 
we will discuss our $\gamma$-ray emission model and  the thickness of the
 outer gap with the pair-creation processes. In section~\ref{result}, 
we will compare the simulated number of the $\gamma$-ray pulsars detected 
on the Earth and its distributions  with the $Fermi$ observations 
(section~3.1 and 3.2), and 
we will predict the future observations (section~3.3). 
In section~\ref{discussion}, we will also discuss ratio of
 the $\gamma$-ray-selected and radio-selected $\gamma$-ray pulsars with
 different radio surveys (section~4.1), and will compare 
 Galactic distributions of the simulated $\gamma$-ray pulsars and 
of the $Fermi$ unidentified sources (section~4.2).  In section~\ref{conclusion},
 we will summarize the results of our population study.
  
\section{Theoretical model and Monte Carlo Simulation}
\label{simulation}
In this section, we will describe our Monte Carlo simulation and
$\gamma$-ray emission model for the canonical pulsars. 
First we perform the Monte Calro simulation
 for the Galactic neutron stars and for the radio emissions from the pulsars, 
which have been   developed  by several authors (e.g. Sturner \& Dermer
1996; Cheng \& Zhang 1998, Gonthier et al. 2002). With this simulation, 
we will be able to obtain a simulated population of the radio pulsars
 consistent with the observations.  Then we will consider the $\gamma$-ray 
emissions from the outer gap accelerator of each  simulated neutron star. 

We note that we will sample and discuss the $\gamma$-ray emissions from 
the simulated neutron stars  with a characteristic
 age younger ($\tau\equiv P/2\dot{P}$) than 10~Myr. 
This is because the present  known $\gamma$-ray emitting canonical pulsars 
have their characteristic  spin down age  less than 5~Myr.  
Also, we will find that the present $\gamma$-ray emission model and the 
simulation predict that the canonical pulsars with age older than 10~Myr 
are hard to produce the observable $\gamma$-ray emissions in their 
 magnetosphere. 

\subsection{Period and magnetic field} 
 We assume that the neutron stars  are born at a rate of 1$\sim$2 per
 century. The initial period of the new born neutron
 star is not well predicted by theory and observation, except for  several 
cases.  For the Crab pulsar, the initial period is  estimated at 
$P_0\sim 19$~ms. Such short birth rotation 
period is also  expected by the existence of young 16~ms pulsar 
PSR J0537-06910 (Marshall et al. 1998). Although 
most of the new pulsars may have 
initial period $P_0\sim 20$~ms, 
 there is good evidence that some pulsars were 
born with a longer initial period (e.g. $P_0\sim 62$~ms of PSR J1811-1925,
 Kaspi et. al., 2001). 
However, we found  that the distribution of initial rotation 
 period does not affect 
much to the following results of the simulation (e.g. the distribution 
of the radio pulsars and of $\gamma$-ray pulsars). For example, we compared 
the results of two simulations with different 
 distribution of  the initial period; one is that all simulated pulsars are 
randomly distributed in the range  $P_0=20-30$~ms, and other is that 
 70~\% of pulsars are distributed at $P_0=20-30~$ms and 
 30~\% are  $P_0=20-100$~ms. As we will describe in later, 
because the pulsars younger than 5~My can become the $\gamma$-ray 
emitting pulsars, 
we will sample the pulsars younger than 10~My. For the population of the 
relatively younger pulsars, we could see that  
the populations of two simulations are  very similar to each other. 
The difference 
becomes significant if we assume  $>50$~\% are randomly distributed in  
$P_0=20-100$~ms. In this paper, therefore, because we do not 
know well  exact distribution of initial rotation period, we randomly 
choose the initial period in the narrow range $P_0=20-30$~ms,
 such like the assumption  in the previous Monte-Carlo studies.

We assume a Gaussian distribution in $\log_{10} B_s$ for 
 the initial strength of the  magnetic  field on the stellar surface, 
\begin{equation}
\rho_B(\log_{10}B_s)=\frac{1}{\sqrt{2\pi}\sigma_B}\exp
\left[-\frac{1}{2}\left(\frac{\log_{10} B_s-\log_{10}
		   B_0}{\sigma_B}\right)^2\right],
\end{equation}
In this study we apply 
 $\log_{10} B_0=12.6$ and $\sigma_B=0.1$, which produce a
 consistent distribution of the magnetic field of the known radio pulsars 
derived from the dipole radiation model (Manchester et al. 2005). 
 Because we simulate the neutron 
stars with a spin down age younger than $\tau\le 10$~Myrs,  we do not consider 
the evolution of the magnetic field of each simulated neutron star. The
 evolution of the stellar  magnetic field may be important for the neutron 
star with an age larger than $\tau \ge 10$~Myrs 
(Goldreich \& Reisenegger 1992; Hoyos, Reisenegger \& Valdivia 2008).

With a constant stellar magnetic field with time, the period evolves as 
\begin{equation}
P(t)=\left(P_0^2+\frac{16\pi^2R_{s}^6B^2}{3Ic^3} t\right)^{1/2},
\end{equation}
where we assume the pure dipole radiation, $R_s$ is the stellar radius and $I$ is the neutron star 
momentum of inertia. We apply $R_s=10^6$~cm and
 $I=10^{45}~\mathrm{g\cdot cm^2}$ in this study. The time derivative of
 the rotation period is calculated from 
\begin{equation}
\dot{P}(t)=\frac{8\pi^2R_{s}^6B^2}{3Ic^3P}.
\end{equation}

\subsection{Initial spatial and velocity  distributions }
Following the studies done by Paczynski (1990) and Sturner \& Dermer
(1996), we assume the birth location  of the neutron star is described by 
the following distributions,   
\[
\rho_R(R)=\frac{a_{R}\mathrm{e}^{-R/R_{\mathrm{exp}}}R}{R^2_{\mathrm{exp}}},
\]
\begin{equation}
\rho_z(z)=\frac{1}{z_{\mathrm{exp}}}\mathrm{e}^{-|z|/z_{\mathrm{exp}}},
\end{equation}
where $R$ is the axial distance from the axis through the Galactic
center  perpendicular to the Galactic disk and  $z$ is the distance from
the Galactic disk. In addition, $a_R=20$~kpc, 
$R_{\mathrm{exp}}=4.5$~kpc, 
$a_R=[1-\mathrm{e}^{-R_{max}/R_{exp}}(1+R_{max}/R_{exp})]^{-1}$ with $R_{max}=20$
~kpc and $z_{exp}=75$~pc.

For the distribution of the initial velocity of simulated neutron star, 
we refer the result of the study for the radio pulsars done by 
Hobbs et al. (2005).  They found that the distribution of the
three-dimensional velocity of their sample is well described by  
  a Maxwellian
distribution  with a characteristic width of $\sigma_v=265$~km/s, namely,
\begin{equation}
\rho_v(v)=\sqrt{\frac{\pi}{2}}\frac{v^2}{\sigma_v^3}
\mathrm{e}^{-v^2/2\sigma_v^2}.
\end{equation}
We apply this  form  for the distribution of the initial kick 
 velocity due to the super nova explosion.
In this study, we randomly choose  the direction of this kick velocity 
 of the neutron star in the three-dimensional space. For the azimuthal
 component of the velocity,  we add the  circular
 velocity due to the Galactic gravitational 
potential field at the birth position of the neutron star.
The circular velocity  due to the Galactic potential is calculated from 
\begin{equation}
v_{circ}=\left[R\left(\frac{\partial\Phi_{sph}}{\partial R}+
\frac{\partial\Phi_{dis}}{\partial R}+\frac{\partial\Phi_{h}}{\partial R}
\right)\right]^{1/2},
\end{equation}
where $\Phi_{sph}$, $\Phi_{dis}$ and $\Phi_{h}$ are spheroidal, disk
and halo components of the Galactic gravitational potential,  which are
given by Pacynski (1990) as 
\begin{equation}
\Phi_i(R,z)=-\frac{GM_i}{\sqrt{R^2+[a_i+(z^2+b_i^2)^{1/2}]^2}},
\end{equation}
 where $a_{sph}=0$, $b_{sph}=0.277$~kpc and  
$M_{sph}=1.12\times 10^{10}M_{\odot}$ for the spheroidal component,  
$a_{dis}=3.7$~kpc, $b_{dis}=0.20$~kpc, 
and $M_{dis}=8.07\times 10^{10}M_{\odot}$ for the disk component, and  
\begin{equation}
\Phi_{h}(r)=\frac{GM_{c}}{r_c}\left[\frac{1}{2}\ln
 \left(1+\frac{r^2}{r_c^2}\right)+\frac{r_c}{r}\tan^{-1}
\left(\frac{r}{r_c}\right)\right], 
\end{equation}
where $r_c=6.0$~kpc $M_c=5.0\times 10^{10}M_{\odot}$ for the halo
component. 

\subsection{Equation of motion}
To obtain  current position of each simulated neutron star, we 
solve the  equation of motion from its birth to the current time. 
The equation of motion is given by  
\begin{equation}
 \frac{dR^2}{dt^2}=\frac{v_{\phi}^2}{R}-\frac{\partial
 \Phi_{tot}}{\partial R},
\label{eqr}
\end{equation}

\begin{equation}
 \frac{dz^2}{dt^2}=-\frac{\partial \Phi_{tot}}{\partial z},
\label{eqz}
\end{equation}
and 
\begin{equation}
Rv_{\phi}=\mathrm{constant},
\end{equation}
where $\Phi_{tot}=\Phi_{sph}+\Phi_{dis}+\Phi_h$, and $v_{\phi}$ is the
azimuthal component of the velocity. The Lagrangian in units of energy
per unit mass is described by 
\begin{equation}
L=\frac{v^2(R,z,\phi)}{2}-\Phi_{tot}(R,z)
\end{equation}
which is used for maintaining an accuracy of the integration of the
trajectory described by equations~(\ref{eqr}) and (\ref{eqz}).
We maintain the accuracy of one part per $10^5$ during the trajectory
from its birth  to the current time. 

\subsection{Radio emission and detection}
Because detail process of the  radio emission 
in the pulsar magnetosphere has not been understood well, we 
apply an empirical relation among  the radio luminosity,  
the rotation period and the period time derivative 
(Cheng \& Zhang 1998; Gonthier et al. 2002). 
We assume that the radio luminosity at 400~MHz, which is defined by 
$L_{400}\equiv d^2S_{400}$ with $d$ being the distance and 
$S_{400}$ observed flux, for 
 each simulated neutron star  follows  the  distribution 
(Narayan \& Ostriker 1990)
\begin{equation}
\rho_{L_{400}}=0.5\lambda^2\mathrm{e}^{\lambda},
\end{equation}
where  $\lambda=3.6[\mathrm{log_{10}}(L_{400}/<L_{400}>)+1.8]$ with
$\mathrm{log}<L_{400}>=6.64+\frac{1}{3}\mathrm{log_{10}}(\dot{P}/P^3)$, and  
$L_{400}$ is the luminosity in units of  $\mathrm{mJy~kpc^2}$.  
The radio flux on the
Earth is given by $S_{400}=L_{400}/d^2$.  We scale the simulated 
400~MHz luminosity to the observation frequency of each survey using 
a typical photon index -2.

The radio emissions from the pulsars may not point toward the Earth. 
We apply the radio beaming fraction described as  (Emmering \& Chevalier 1989) 
\begin{equation}
f_r(\omega)=(1-\cos\omega)+(\pi/2-\omega)+\sin\omega,
\end{equation}
where $\omega$ is the half-angle (in radian) of the radio emission
cone and $0\le f_r \le 1$. We apply the half-angle of the radio emission
cone  studied  by Kijak \& Gil (1998, 2003), 
\begin{equation}
\omega_{KG}\sim 0.02r^{1/2}_{KG}P^{-1/2}, 
\end{equation}
 where  
\[
r_{KG}=40\nu^{-0.26}_{GHz}\dot{P}^{0.07}_{-15}P^{0.3},
\]
where $\dot{P}_{-15}$ is the period time derivative in units of
$10^{-15}$ and $\nu_{GHz}$ is the radio frequency in units of GHz.  
In the Monte-Carlo simulation, the fraction
$f_r$ of the pulsars with same width $\omega$ can emit the radio beam 
toward the Earth. 

The minimum detectable radio flux (that is, sensitivity)
 of a particular radio survey is usually express as (Edwards et al. 2001), 
\begin{equation}
S_{min}=\frac{C_{thres}[T_{rec}+T_{sky}(l,b)]}{G\sqrt{2B_{BD}t_i}}
\sqrt{\frac{W}{P-W}},
\label{smin}
\end{equation}
where $C_{thres}$ is the receiver detection threshold S/N, $T_{rec}$ is the
receiver temperature, $T_{sky}(l,b)$ is the sky temperature with $l$
being Galactic longitude and $b$ Galactic latitude, $G$ is the telescope gain,
$B_{BD}$ is the band width, $t_i$ is the integration time and $W$ is the
observed pulse width.  The sky
temperature at the frequency $\nu$  
is given by (Johnston et al. 1992) 
\begin{equation}
T_{sky}(\nu)=25+\left\{\frac{275}{[1+(l/42)^2)][1+(b/3)^2]}\right\}
\left(\frac{408~\mathrm{MHz}}{\nu}\right)^{2/6}~\mathrm{K}. 
\end{equation}
The observed pulse width  is given by 
\begin{equation}
W^2=W_0^2+\tau_{samp}^2+\tau_{DM}^2+\tau_{scat}^2,
\end{equation}
where $W_0\sim 0.05P$ is the intrinsic pulse width, $\tau_{samp}$ is the
receiver sampling time, $\tau_{DM}$ is the frequency dispersion of the
radio pulse as it passes the interstellar medium, and $\tau_{scat}$ is
the frequency dispersion due to the interstellar scattering. The
dispersion $\tau_{DM}$ has the form 
\begin{equation}
\tau_{DM}=8.3\times 10^6\mathrm{DM} \left(\frac{\delta\nu}{\mathrm{MHz}}
\right)\left(\frac{\nu}{\mathrm{MHz}}\right)^{-3}~\mathrm{ms} 
\end{equation}
where $\delta\nu$ is the frequency channel bandwidth of the receiver 
and DM (in units of
 $\mathrm{pc~cm^{-3}}$)
 is the dispersion measure, which is calculated from 
$\mathrm{DM}=\int n_e(R,Z) ds$, where $s$ is the distance to the source
 and $n_e(R,Z)$ is electron number density. The Galactic distribution
 of the electron number density is calculated from the model investigated by 
Cordes \& Lazio (2002). The dispersion due to the interstellar
scattering is obtained by 
\begin{equation}
\tau_{scat}(\nu)=\tau_{scat,400}\left(\frac{400~\mathrm{MHz}}{\nu}\right)^{4.4}
~\mathrm{ms}
\label{tauscat}
\end{equation}
where $\tau_{scat,400}=10^{-4.61+1.14\mathrm{log_{10}~DM}}
+10^{-9.22+4.46\mathrm{log_{10}~DM}}$.

\begin{figure}
\epsscale{.80}
\plotone{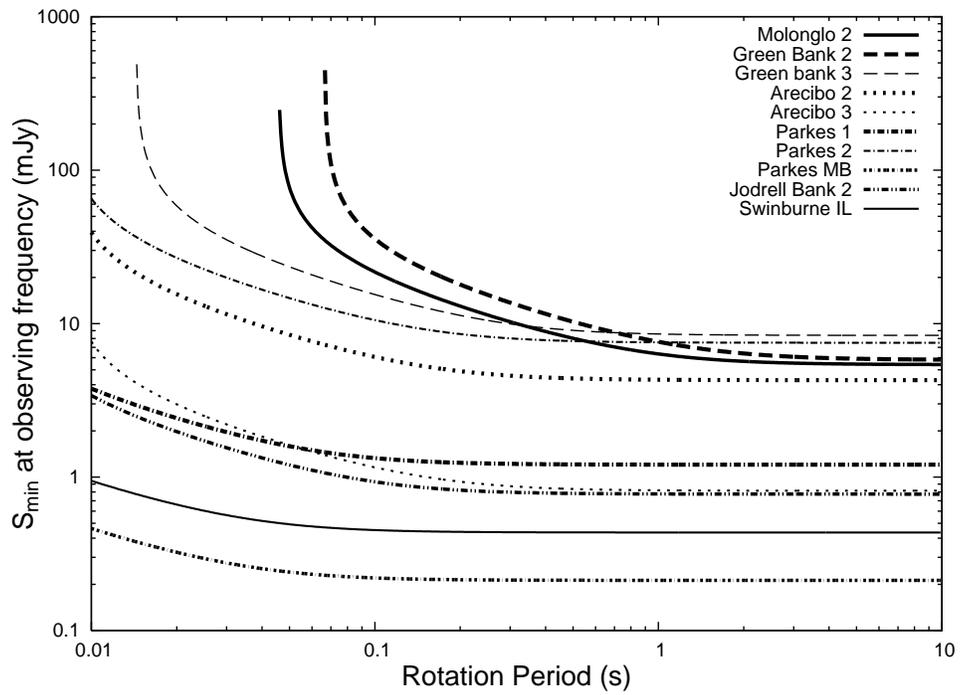}
\caption{Radio flux threshold for each radio survey as a function of 
the rotation period. The results are for $T_{sky}=150$~K at 408~MHz and the dispersion measure of $DM=200~\mathrm{cm^{-3}pc}$.}
\label{sensitive}
\end{figure}

In Table~1, we list the radio surveys that we applied in the present 
simulation and its receiver parameters 
in equations~(\ref{smin})-(\ref{tauscat}). Also, Figure~\ref{sensitive}
plots the sensitivity described by equation~(\ref{smin}) for each pulsar survey  
as a function of the period. In Figure~\ref{sensitive}, we used the sky
temperature of $T_{sky}=150$~K at 408~MHz and the dispersion measure of 
DM=200~$\mathrm{cm^{-3}}$pc. We note that although 
new radio pulsars have been reported with the search of the pulsed emissions 
for the specific sources (e.g.  Camilo et al 2009  for $Fermi$ sources), 
the  majority of the present radio pulsars 
have been discovered by the surveys listed in Table~1. Therefore 
we do not take into account such pulse  search for the individual sources.

\subsection{$\gamma$-ray emission model}
\label{gemission}
In this section, we describe our  $\gamma$-ray emission model applied
for each simulated pulsar. We consider the $\gamma$-ray emission from the 
outer gap accelerator (Cheng, Ho \& Ruderman 1986a,b; Zhang \& Cheng 1997;
Takata et al. 2010).  In the outer gap, the charged particles 
(electrons and positrons) are accelerated up to 
a Lorentz factor of $\Gamma\ge 10^7$ by the electric field along the magnetic field line. 
The accelerated particles can emit several GeV $\gamma$-ray photons through 
the curvature radiation process. Assuming the force balance between the 
electric force and the radiation drag force, the Lorentz factor is described by 
\begin{equation}
\Gamma=\left(\frac{3R_c^2}{2e}E_{||}\right)^{1/4},
\end{equation}
where $R_c$ is the curvature radius of the magnetic field lines and $E_{||}$ is
 the accelerating electric field. The accelerating electric 
field in the outer gap is approximately described 
by (Cheng et al. 1986a,b; Cheng, Ruderman \& Zhang 2000)
\begin{equation}
E_{||}(r)\sim \frac{B(r)f_{gap}(r)^2R_{lc}}{R_c},
\end{equation}
where $R_{lc}=cP/2\pi$ is the light radius and $f_{gap}(r)$ is so called 
fractional gap thickness in the poloidal plane. 
The power of the curvature radiation emitted by the individual 
particle is written as 
\begin{equation}
P_c(E_{\gamma},r)=\frac{\sqrt{3}e^2\Gamma}{hR_c}F(x),
\end{equation}
where $x=E_{\gamma}/E_c$ with $E_c=3hc\Gamma^3/4\pi R_c$ and 
\[
F(x)=x\int_x^{\infty}K_{5/3}(t)dt,
\]
where $K_{5/3}$ is the modified Bessel function of the order 5/3.  
If the $\gamma$-ray beam  points toward an observer, the observer  
will measure the phase-averaged spectrum of
 (Hirotani 2008), 
\begin{equation}
\frac{dF_{\gamma}}{dE_{\gamma}}\sim N(r_{\xi}) R_c(r_{\xi}) P_c(E_{\gamma},r_{\xi})\frac{\delta A}{d^2},
\label{flux}
\end{equation}
where $N$ is the typical particle number density, $r_{\xi}$ represents the 
typical radius at emission 
positions from which the emissions can be measured by the observer
 and   $d$ is the distance to the source.
 In addition,  $\delta A$ is 
the cross section of the gap 
perpendicular to the magnetic field lines and 
 is estimated from $\delta A\sim [B_s/B(r_{\xi})]R_p^2\delta\theta \delta\phi$, 
where $R_p$ is the polar cap radius, and $\delta\theta$ and $\delta\phi$ is 
gap width in the colatitude  and azimuthal directions measured on the 
stellar surface, respectively.
 For the typical case of the outer gap model,
 we apply the Goldreich-Julian value, $N=B/Pce$, for the number density, 
and the curvature radius of $R_c=R_{lc}/2$. We assume that the outer gap is 
typically extending to the azimuthal direction with $\delta\phi\sim \pi$. 
The integrated energy flux 
between 100~MeV and 300GeV can be calculated from
\begin{equation}
F_{\gamma,>100~MeV}=\int_{100MeV}^{300GeV}\frac{dF_{\gamma}}{dE_{\gamma}}dE_{\gamma}.
\end{equation}

 In the paper, we assume  that the gap current is order of 
Goldreich-Julian value over the full width  as  zeroth order approximation. 
 The detailed calculation 
of the outer gap model (e.g. Takata \& Chang 2007) shows that 
the pair-creation process produces the distribution of the current density  
 in the direction perpendicular to the magnetic field lines. 
On the other hand, recent two-layer outer gap model (Wang et al. 2010) predicts 
that total current in the gap is order of the Goldreich-Julian 
value. Therefore we expect that as long as we discuss 
the total power of the $\gamma$-ray radiation 
from the outer gap accelerator, the uniform current distribution with 
the Goldreich-Julian value  
 would not be bad approximation. The distribution of the current in the 
gap will be more  important to discuss 
the spectral shape above 100~MeV bands. 

The beaming  fraction $f_{\gamma}$, which is defined by ratio 
of the solid angle of the $\gamma$-ray beam to $4\pi$,
 affects number of the detected $\gamma$-ray pulsars. However, 
it is complicated to describe the beaming fraction with 
a simple function. This is because the beaming  fraction should be 
different for different inclination angle and for different pulsars. 
For the outer gap model, however, the most of emissions  concentrate within 
 the direction   $90^{\circ}\pm 35^{\circ}$  measured from the rotation axis, 
as we can see in the photon mappings  (e.g. Cheng, Ruderman \& Zhang 2000;
 Takata \& Chang 2007), 
indicating  the beaming fraction  is roughly  
$f_{\gamma}\sim35^{\circ}/90^{\circ}\sim 0.4$. 

Let's discuss the fractional gap thickness $f_{gap}$, which determines 
 the $\gamma$-ray flux of equation~(\ref{flux}). 
Recently, Wang et al.  (2010) studied the relation between the 
$\gamma$-ray luminosity and the spin down luminosity for the $\gamma$-ray 
pulsars detected by the $Fermi$ telescope. They found that the $\gamma$-ray 
luminosity  can be described by
\begin{eqnarray}
L_{\gamma}\propto L_{sd}^{\beta}\left\{
\begin{array}{ll}
\beta\sim 0,& \mathrm{for}~L_{sd}\gtrsim 10^{36}~\mathrm{erg/s}\\
\beta\sim 0.5,& \mathrm{for}~L_{sd}\lesssim 10^{36}~\mathrm{erg/s}.
\end{array}
\right.
\label{relation}
\end{eqnarray} 
 They further argued that because the typical $\gamma$-ray luminosity is 
described by $L_{\gamma}\sim f_{gap}^3L_{sd}$, 
the change of the relation
 between $L_{\gamma}$ and $L_{sd}$ indicates  
the gap closure mechanism in the poloidal plane switches  
 at $L_{sd}\sim 10^{36}$~erg/s.

The relation that  $L_{\gamma}\propto L_{sd}^{\beta}$ with $\beta\sim 0$ in 
equation~(\ref{relation}) may be understood with the gap closure mechanism 
proposed by Zhang \& Cheng (1997). They argued that the photon-photon 
pair-creation process between the $\gamma$-rays emitted in the outer gap and the
 X-rays from the stellar surface controls the gap activities. 
 They discussed that the
X-rays from  the heated polar cap by the return particles  will collide
 with $\gamma$-rays in the outer gap and provide 
 a self-consistent mechanism to restrict the fractional size of
the gap. They estimated the typical gap size near the light cylinder as 
\begin{equation}
f_{ZC}=5.5P^{26/21}B_{12}^{-4/7},
\label{fzc}
\end{equation}
where $B_{12}$ is the polar magnetic field strength in units of $10^{12}$~G. 
With this model,  the $\gamma$-ray luminosity and 
the spin down power is related as 
\begin{equation}
L_{\gamma}\sim 1.3\times 10^{34}B^{1/7}_{12}L_{sd,36}^{1/14}~\mathrm{erg/s},
\end{equation}
where we used 
$L_{sd,36}$ is the spin down power in units of $10^{36}$~erg/s.

On the other hand, the relation that 
 $L_{\gamma}\propto L_{sd}^{\beta}$ with $\beta\sim 0.5$ 
in equation~(\ref{relation}) may be explained by the gap closure model  
proposed by Takata et al. (2010), in which the outer gap is closed by  the
magnetic pair-creation process near the stellar surface. 
They argued that the returning particles, which were
produced by the photon-photon pair-creation process in the gap, will
emit the  photons with an energy  $m_ec^2/\alpha_f\sim 100 \rm
MeV$ by the curvature radiation near the stellar surface.  
They discussed  that those 100MeV photons can  become pairs by magnetic
pair creation process and that  
these secondary pairs can continue to radiate several MeV photons via 
 the synchrotron radiation. The photon multiplicity is easily
over $10^5$ per each incoming particle.  For a simple dipole field
structure, all  pairs should move inward and  cannot
affect to the outer gap accelerator.
 However they argued that the existence of strong 
surface local field (e.g. Ruderman 1991, Arons 1993) has been widely
suggested. In particular if the field lines near the surface, instead of
nearly perpendicular to the surface, are bending sideward due to the
strong local field,  the pairs created in these local magnetic field
lines can have an angle bigger than 90$^{\circ}$, which results in an
outgoing flow of pairs. In fact it only needs a very tiny fraction
(1-10) out of $10^5$ photons creating pairs in these field lines, which
are sufficient to provide screening in the outer gap when they migrate
to the outer magnetosphere. With this model, they estimated  
the fractional gap thickness when this situation occurs as
\begin{equation}
f_m= 0.8 K  P^{1/2},
\label{fm}
\end{equation}
where  $K\sim B_{m,12}^{-2}s_7$ is the parameter characterizing  the
local parameters, $B_{m,12}$ and $s_7$ are the local magnetic field 
in units of $10^{12}$G and the local curvature radius in units 
of $10^7$cm, respectively. They argued that $K\sim 2$ for the canonical 
pulsars, while $K\sim 15$ for the millisecond pulsars. 
This gap closure model predicts that the $\gamma$-ray luminosity is related 
with the spin down power as 
\begin{equation}
L_{\gamma}\sim 1.1\times 10^{34}K^{3}B^{3/4}_{12}L^{5/8}_{sd,36}~\mathrm{erg/s}.
\end{equation}

We can see that the fractional gap thickness 
 $f_{ZC}$ is smaller than $f_m$ for the younger pulsar, implying 
the gap thickness is determined by 
the photon-photon pair-creation process, while $f_m<f_{zc}$ for the
older pulsars, implying the magnetic pair-creation process may control
the gap thickness.   Equating $f_{ZC}$ and $f_m$, we obtain the critical
spin down power that 
\begin{equation}
L_{sd,c}\sim 10^{36}K^{-168/31}B_{12}^{-34/31}~\mathrm{erg/s},
\end{equation}
which may explain the switching positions in equation~(\ref{relation})  
argued by Wang et al. (2010). 
In this paper, we will investigate the population predicted 
by  the $\gamma$-ray emission model 
with the switching of the gap closure mechanism at $L_{sd,c}$, because 
the population with the emission model without the switching (that is 
$f_{gap}=f_{zc}$) have been investigated in the previous studies 
(e.g. Chang and  Zhang 1998; Zhang, Zhang and Cheng 2000; Zhang et al. 2004).

Finally we would like to discuss 
the maximum fractional thickness, $f_{max}$, for  
the active outer gap accelerator. 
 Zhang \& Chang (1997) argued that the pulsar with the fractional 
gap thickness larger  than
unity, $f_{ZC}>1$, is not active, because the pairs are not created in the
gap by the photon-photon pair-creation process.  However, 
it may be possible that
the maximum gap thickness is smaller than unity. For example,  outer gap
accelerator can exist between the last-open field lines and the critical
field lines, on which the Goldreich-Julian charge density is  equal to
zero at the light cylinder. Therefore,  we may define the
maximum gap thickness  as 
\begin{equation}
f_{crit}=\frac{\theta_P-\theta_c}{\theta_p},
\end{equation}
where $\theta_p$ and $\theta_c$ are polar angles of the last-closed filed
line and of the critical field line on the stellar surface,
respectively.  For the pure dipole field, we obtain
\begin{equation}
\theta_{p}=\alpha+\sin^{-1}\left[\sin(\theta_{lc}-\alpha)
\left(\frac{R_s}{R_{lc}}\sin\theta_{lc}\right)^{1/2}
\right]
\end{equation}
and 
\begin{equation}
\theta_{c}=\alpha+\sin^{-1}\left[\sin(\theta_{n}-\alpha)
\left(\frac{R_s}{R_{lc}}\sin\theta_{n}\right)^{1/2}
\right ],
\end{equation}
respectively, where $\alpha$ is
the inclination angle, 
\[
 \theta_{lc}=\tan^{-1}\left(\frac{-3-\sqrt{9+8\tan^2\alpha}}{4\tan\alpha}
\right),
\]
and 
\[
\theta_n=\tan^{-1}\left(\frac{3\tan\alpha+\sqrt{8+9\tan^2\alpha}}{2}\right).
\]
In the realistic situation, we expect that 
the polar cap shape and the magnetic field configuration in the pulsar
magnetosphere are more complicated and the simple dipole formulae for
 $\theta_{p}$ and $\theta_c$ may not be applied. In this paper, therefore, 
we examine two extreme cases that $f_{max}$=1 and
 $f_{max}=(\theta_p-\theta_c)/\theta_p$ with the dipole field, where we 
randomly choose  the inclination angle  $\alpha$.   
These results of the two case may  give a range of  uncertainty 
 of the present theoretical predictions.

\section{Results}
\label{result}
\label{Radio population}
In this paper, we sample the pulsars with a characteristic age smaller
than $\tau\le 10~$Myr, because (1) we are interesting in the
population study for the $\gamma$-ray emitting canonical pulsars and 
 (2) the present 
canonical $\gamma$-ray pulsars have been detected 
with a characteristic age younger 
than $5$~Myr. In fact, although 
we run the simulation to  obtain more than  200 times of the present radio
pulsars,  we could not find any detectable canonical $\gamma$-ray pulsars with 
a characteristic age older than $10$~Myr.

\begin{figure}
\epsscale{.80}
\plotone{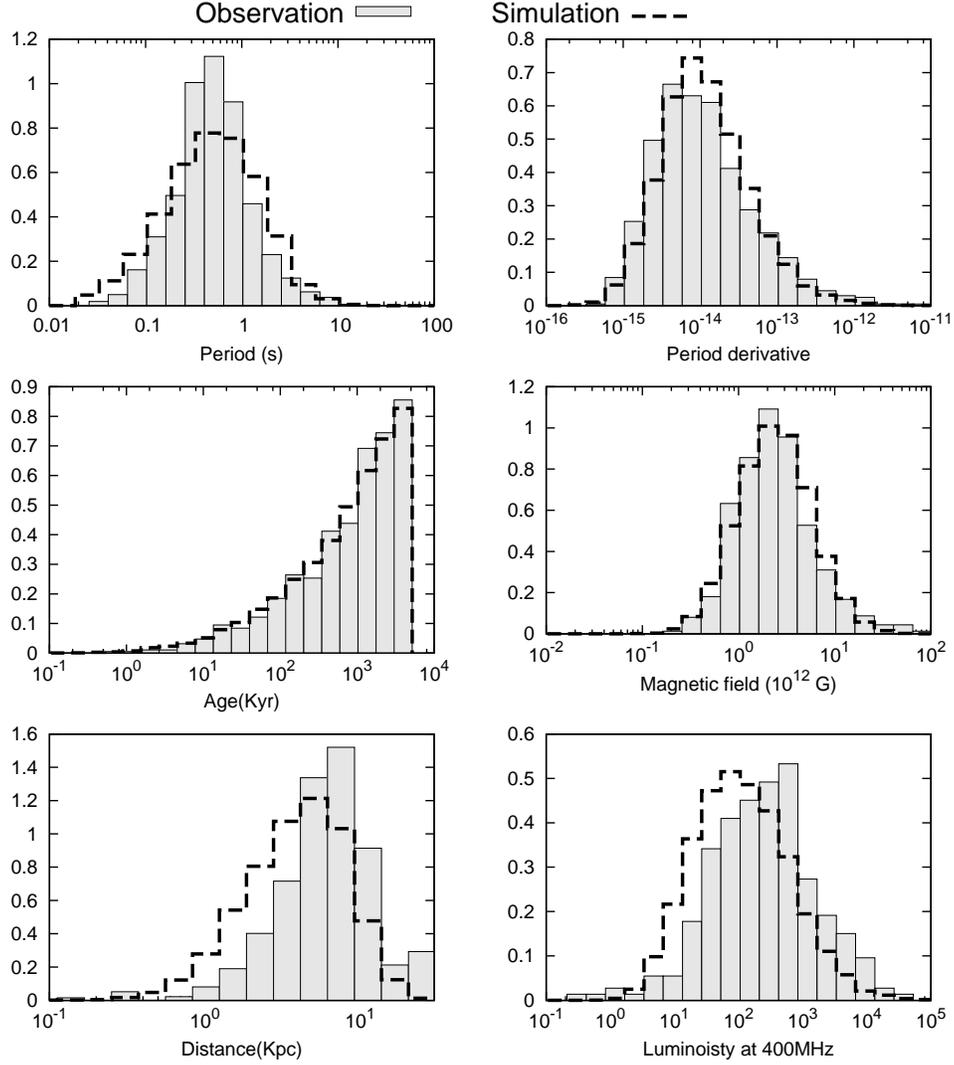}
\caption{Distributions of various  characteristics of the radio pulsars with 
a characteristic age ($\tau=P/2\dot{P}$) younger than 5~Myr. The shaded 
histograms and unshaded histograms present results for the observation and 
simulation, respectively. The data were taken from Manchester et al. (2005).}
\label{disprop-rad}
\end{figure}

We simulate $3\times 10^6$ of neutron 
 stars with a constant birth rate during $10$~Myrs,  
and detect 16602 neutron stars as the radio pulsars with a 
characteristic age of $\tau\le 5$Myr. Scaling the simulated
 population  to  the observation,  $\sim 800$ radio pulsars,
 provides $\sim$1.3 per century as the predicted 
 birth rate. 
 Figure~\ref{disprop-rad} compares the distributions of the various 
characteristics of the  simulated (unshaded histograms) and observed 
pulsars (shaded histograms) with a characteristic age of $\tau\le 5$~Myr. 
We can see in Figure~\ref{disprop-rad}  
that the present simulation  reproduces well
 the observed  distributions for the period time derivative, the spin down 
age and the surface magnetic field, although the simulation
 predicts more pulsars close to
 the Earth and  more pulsars with a smaller radio flux, as bottom two 
panels show. The difference between the results of the 
observation and of the simulation  may be caused 
 by the  assumptions within the present
 simulation.  
On the other hand, we expect that an adjustment of the slight 
difference between the simulated  
and observed population   in 
 Figure~\ref{disprop-rad} will  not change much  the results of population on 
 the $\gamma$-ray  pulsars discussed in the following sections.

\subsection{Population ``bright'' $\gamma$-ray pulsars} 
\label{numberp}
First, we sample only 
bright $\gamma$-ray pulsars,  because the $\gamma$-ray
pulsars with a larger flux measured on the Earth will be
 preferentially discovered by the $Fermi$ $\gamma$-ray telescope.
Moreover, because the detection of the bright point sources 
might not be affected much 
by $\gamma$-ray background emissions,  it  might be true  
that most of (or all) bright $\gamma$-ray pulsars  
have already discovered by the $Fermi$, implying the distribution of the bright 
$\gamma$-ray sources will be able to used to test the emission model.  
 In this paper,  we define  ``bright'' $\gamma$-ray pulsars
with a flux larger than $F_{\gamma}\ge 10^{-10}~\mathrm{erg/cm^2s}$. 
In the first $Fermi$ pulsar catalog (Abdo et al. 2010a), the number of the 
bright $\gamma$-ray pulsars is 12 for the radio selected and 13 for 
the $\gamma$-ray-selected pulsars (including the Gemniga pulsar). Later, 
Saz Parkinson et al. (2010) reported new eight $\gamma$-ray pulsars discovered 
in blind frequency searches. Four out of eight pulsars show the $\gamma$-ray
 emissions brighter than $F_{\gamma}\ge 10^{-10}~\mathrm{erg/cm^2 s}$.

In Table~2, we show the simulated population of
 the radio-selected $N_{r,F>10^{-10}}$ and $\gamma$-ray-selected 
$N_{g,F>10^{-10}}$ bright $\gamma$-ray pulsars. 
 We assumed the beaming factor of the $\gamma$-ray emissions are 
$f_{\gamma}=0.4$ and the birthrate is 1.3 per century.  We present
 the results for $K=1$ in equation~(\ref{fm}), 
the maximum gap thickness  of $f_{max}=f_{crit}$ and of $f_{max}=1$. 

As Table~2 shows, we predict that there are  
 $N_{r,F>10^{-10}}=18\sim23$ radio-selected  and $N_{g,F>10^{-10}}=26\sim34$ 
$\gamma$-ray-selected  bright  $\gamma$-ray pulsars. 
Although  the  ratio of the simulated $\gamma$-ray-selected  
to radio-selected bright
 sources, 1.3-1.5,  is consistent with   
the present $Fermi$ bright pulsars, $17/12\sim 1.4$,
 our simulation predicts more 
number than  the observed number (12 radio-selected and 17 $\gamma$-ray 
selected pulsars).
 To explain this difference, we  may expect that although all bright 
sources are actually discovered 
by the $Fermi$, the detection of the pulsations were failed  due to the
 technical problem, which may be caused by the strong background etc.
 The recent 
$Fermi$ first source catalog  (Abdo et al. 2010c\footnote{$\mathrm{http://fermi.gsfc.nasa.gov/ssc/data/access/lat/1yr_{-}catalog}$})  includes  about 50
 unidentified sources with  a flux $F_{\gamma}\ge
 10^{-10}~\mathrm{erg/cm^2 s}$, 
and most of them can be categorized as a steady source.
The present model indicates that   $15\sim 28$ of the unidentified 
$Fermi$ bright sources  originate  from the canonical $\gamma$-ray pulsars.

\begin{figure}
\epsscale{.80}
\plotone{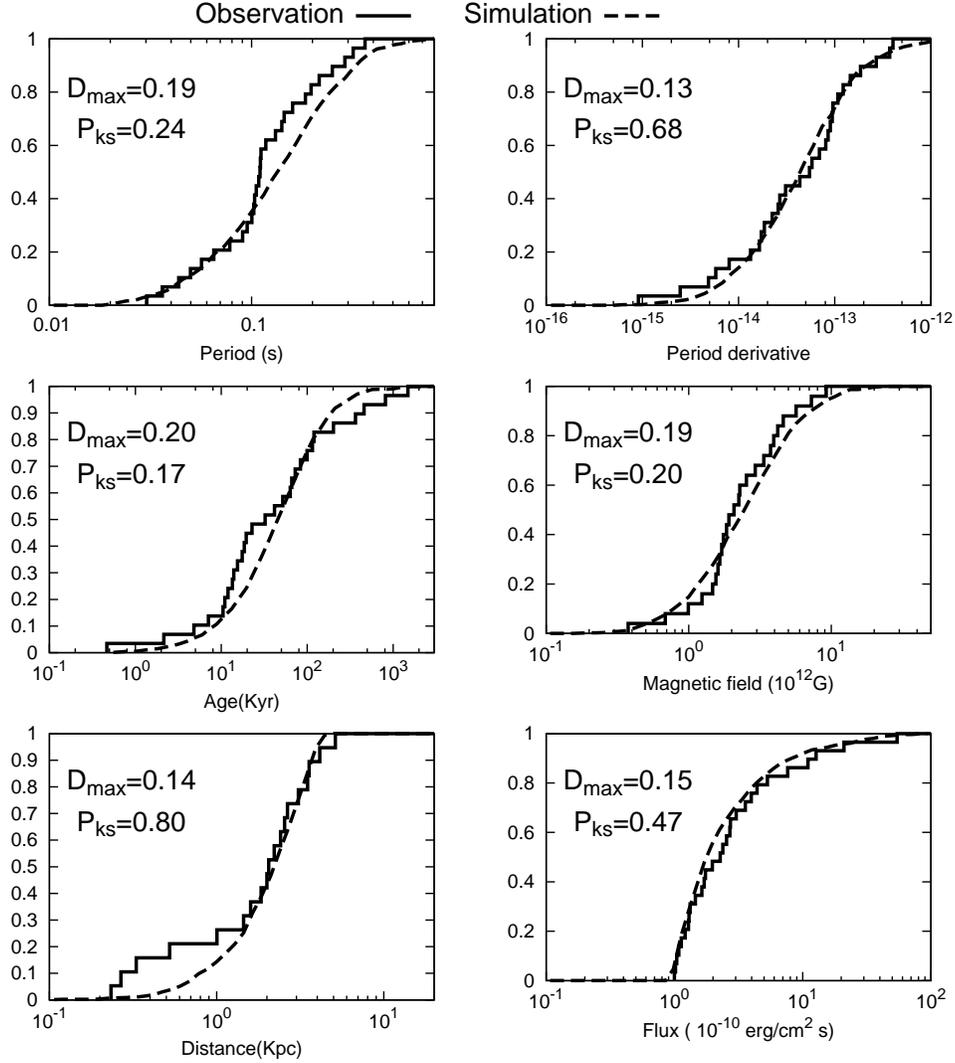}
\caption{Comparison of the cumulative distributions for the various characteristics of 
"bright" $\gamma$-ray pulsars, $F_{\gamma}\ge\ 10^{-10}~\mathrm{erg/cm^2 s}$, including both radio-selected and 
$\gamma$-ray-selected $\gamma$-ray pulsars. The solid and dashed lines represent the distributions for the 
observations and for the simulation, respectively. The p-value ($P_{KS}$) of the Kolmoglov-Smirov test 
and the maximum difference ($D_{max}$) between two distributions are also indicated in 
each panel. The data were taken from Abdo et al. (2010a) and 
Saz~Parkinson et al. (2010). }
\label{distri1}
\end{figure}

 Figure~\ref{distri1} compares  cumulative distributions of
 various characteristics for the simulated and observed 
bright $\gamma$-ray pulsars, 
including the radio-selected and $\gamma$-ray-selected $\gamma$-ray pulsars.  
The results of the simulation are for $f_{max}=f_{crit}$. 
We discuss the combined cumulative 
distribution of the radio-selected  and $\gamma$-ray-selected pulsars,
 (1) because we  increase the  number of  observation sample (12+17=29) 
to increase the statistical accuracy,  and  
 (2) because  some $\gamma$-ray-selected pulsars have been identified 
in radio bands (Camilo et al. 2009).

We performed a Kolmogorov-Smirnov (KS) test to compare the two cumulative 
distributions for the bright sources. In Figure~\ref{distri1}, we present 
 the maximum deviation ($D_{max}$) between the two 
distributions and the p-value ($P_{ks}$) of the KS-test. 
 It can be seen that the  hypothesis that the two distributions are 
drawn from the same parent population  can not be rejected at 
better than $\sim 80$~\% confidence level for the period, spin-down age 
and the surface magnetic field, than $\sim50$~\% for the period time 
derivative and  the flux, and than $\sim 20$~\% for the distance, 
 indicating  those  model distributions
 will be consistent with the observations.  

\subsection{Comparison with $Fermi$ $\gamma$-ray pulsars}

In this section,  we present  simulated population of   the $\gamma$-ray pulsars
 applying the $Fermi$ sensitivities of the six-month observations, and compare 
the simulated population with all known canonical $\gamma$-ray pulsars 
listed in Abdo et al. (2010a) and  Saz~Parkinson et al. (2010).
For the radio-selected $\gamma$-ray pulsars, 
the Fermi sensitivities of the six-month observation 
 is approximately described by 
$F=2\times 10^{-8}~\mathrm{photons/cm^2 s}$ for the Galactic latitude 
$|b|\ge 5^{\circ}$ and $6\times 10^{-8}~\mathrm{photons/cm^2 s}$ for $|b|\le 5^{\circ}$,   
while for the $\gamma$-ray-selected pulsars, 
$F=4\times 10^{-8}~\mathrm{photons/cm^2 s}$ for $|b|\ge 5^{\circ}$ 
and $1.2\times 10^{-7}~\mathrm{photons/cm^2 s}$ for $|b|\le 5^{\circ}$. 

\begin{figure}
\epsscale{.80}
\plotone{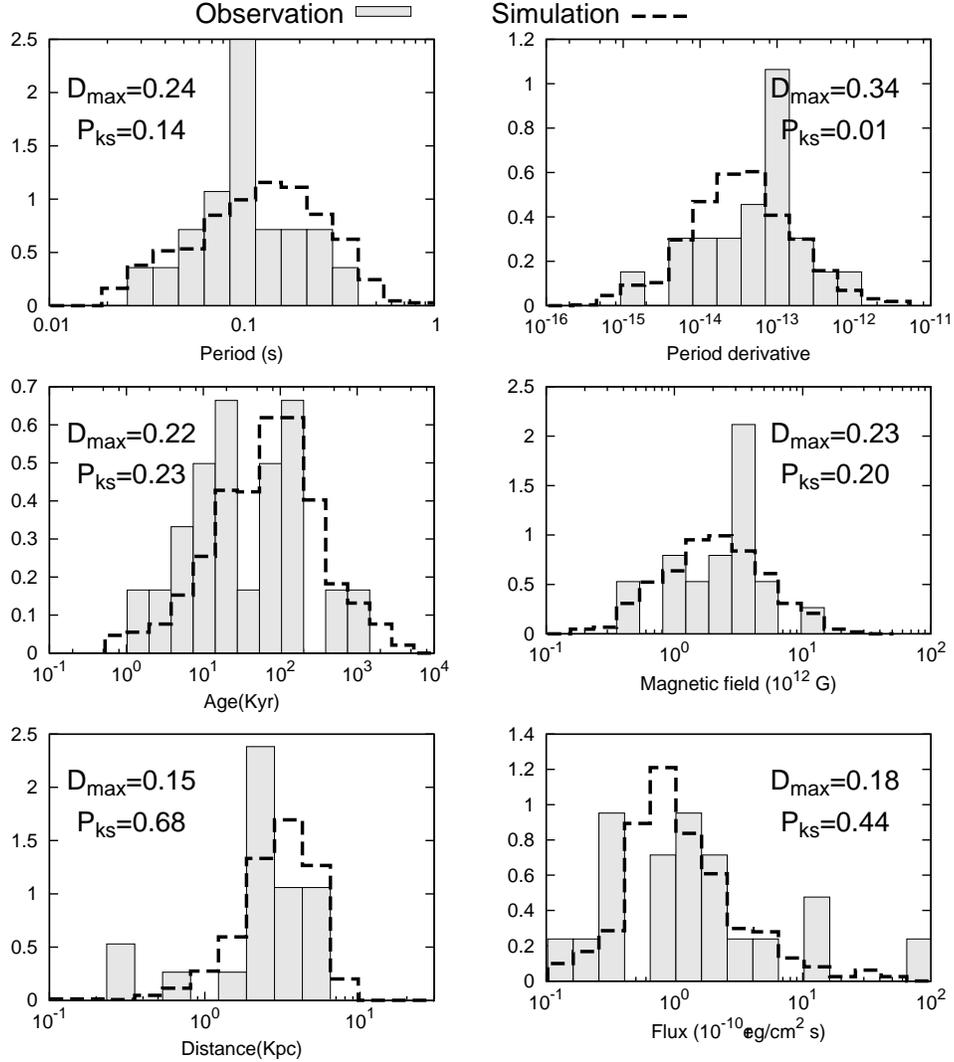}
\caption{The distribution of the various characteristics for 
the radio-selected $\gamma$-ray pulsars for six-month $Fermi$ observations.
 The shaded and unshaded histograms  presents results 
for the observations and for the  simulation, respectively. In addition, $D_{max}$ and $P_{KS}$ show 
the maximum difference of the cumulative distributions  and the p-value of the KS-test, respectively. In the simulation, the sensitivity of the 
$\gamma$-ray detection is set at $F=2\times 10^{-8}~\mathrm{photons/cm^2 s}$ for $|b|\ge 5^{\circ}$ and $6\times 10^{-8}~\mathrm{photons/cm^2 s}$ for $|b|\le 5^{\circ}$. 
The data were taken from Abdo et al. (2010a) and Saz~Parkinson et al. (2010). }
\label{gamma-rl}
\end{figure}

\begin{figure}
\epsscale{.80}
\plotone{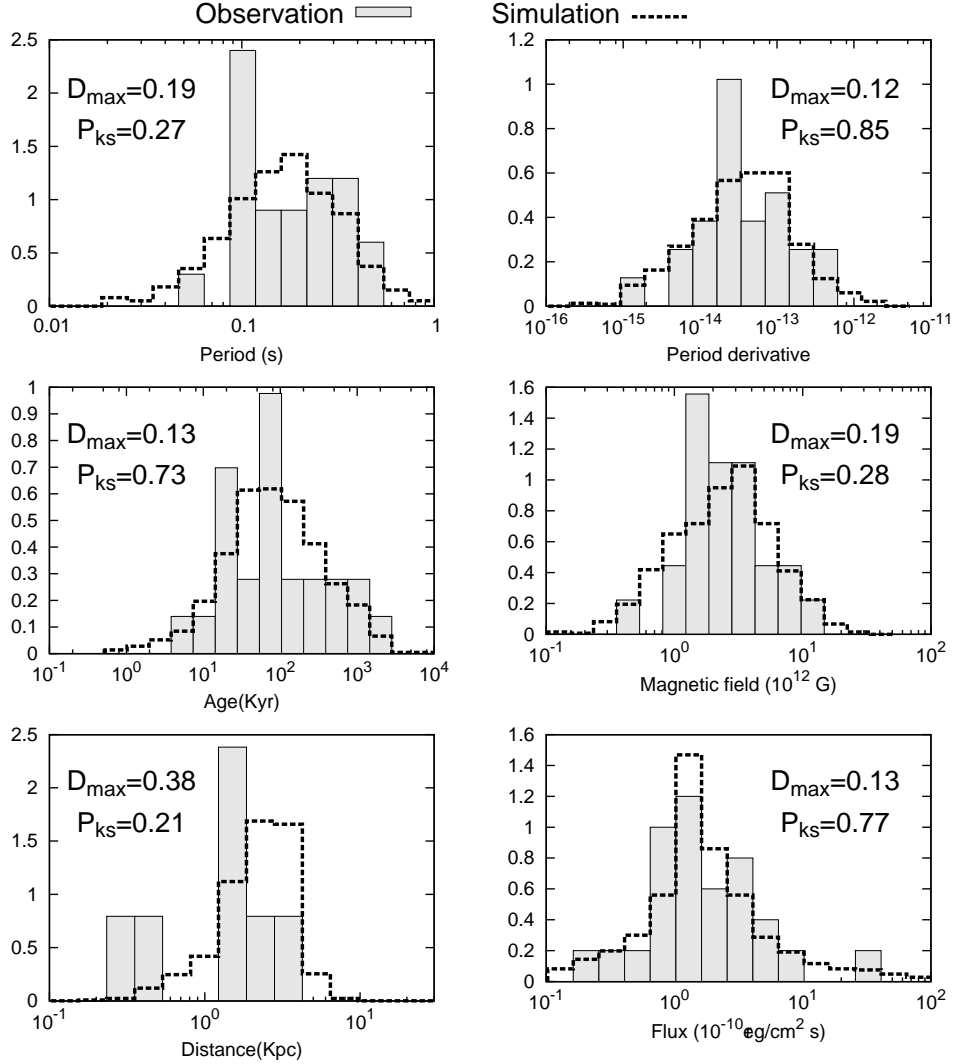}
\caption{The same with Fig.~\ref{gamma-rl}, but results for 
the $\gamma$-ray-selected $\gamma$-ray pulsars.  In the simulation, the sensitivity of 
the $\gamma$-ray detection is set at $F=4\times 10^{-8}~\mathrm{photons/cm^2 s}$ for $|b|\ge 5^{\circ}$ and $1.2\times 10^{-7}~\mathrm{photons/cm^2 s}$ for $|b|\le 5^{\circ}$.}
\label{gamma-rq}
\end{figure}

Figures~\ref{gamma-rl} and~\ref{gamma-rq} show the distributions for 
the various characteristics for the radio-selected and $\gamma$-ray-selected 
$\gamma$-ray pulsars, respectively. The shaded and un-shaded histogram show 
the results for the $Fermi$ pulsars  and  simulated pulsars, respectively.
 The maximum deviation of the cumulative distributions ($D_{max}$) 
 and the p-value of the K-S test ($P_{ks}$) are also indicated in each panel. 
The results are for $f_{max}=f_{crit}$. 
We can see that except for the distributions of the period time derivative 
of the radio-selected $\gamma$-ray pulsars, $p$-vale of the K-S test is larger 
than $P_{ks}\sim 0.15$, indicating hypothesis that the two distributions are 
drawn from the same parent population  can not be rejected at 
better than $\sim 85$~\%. 
Therefore, we would conclude that distributions of the our simulation 
are not conflict with the $Fermi$ observations. 

Although KS-statistic indicates the consistency between the simulated and the 
observed distributions, one can see an excess in the observed distribution 
(shaded histograms in Figures~\ref{gamma-rl} and~\ref{gamma-rq}) compared 
with the simulated distribution at rotation period 
$P\sim 0.1$~s, the  spin down age $\tau\sim 10^5$~yr and the distance 
$d\sim 1$~kpc. This may be cause by intrinsic feature of the pulsars 
or  selection effect or enhancement of the birth rate due to the local 
Galactic structure. Because the excess is caused by only three or four 
pulsars, all possibilities can be considered.
If the excess of the pulsars is
 due to the local structure, the Gould 
Belt may be possible site to produce 
the pulsars which make the excess in the distributions. For example, 
Cheng et al. (2004) discussed that about 20 $\gamma$-ray pulsars with a flux 
$F_{\gamma}\ge 7\times 10^{-11}$ are associated with the Gould Belt. 
The Gould Belt has an ellipsoidal ring with semi-major and semi-minor 
axes equal to $\sim$0.5 and 0.35~kpc, respectively (Guillout et al. 1998),
 although  a smaller size is also predicted  (c.f. Perrot \& Grenier 2002).
 The distance to the center of 
the Gould Belt from the Sun is $\sim 0.1$kpc, 
indicating the neutron stars are born about 0.25-0.6~kpc away from the Sun. 
If we use the typical velocity of the neutron star $\sim$300~km/s, 
the displacement during $\tau\sim 10^5$~yr is $d\sim 300~\mathrm{km/s}
\times 10^{5}~\mathrm{yr}\sim 0.1$kpc. Such a small 
displacement will  not be able to explain 
 the excess  of the distance at 1~kpc, although  
 the size of the Gould Belt can play an important role. 
We expect that if the excess in the distributions are caused by the 
selection effects or by the effect of Gould Belt, the excess will decrease 
as increase of the $\gamma$-ray pulsar population. On the other hand, 
 if it is the  intrinsic feature of the pulsar population,
 the excess will remain even the population of the pulsars increase.

In Table~2, we show the number of the simulated radio-selected ($N_{rtot}$) and 
$\gamma$-ray-selected ($N_{gtot}$) $\gamma$-ray pulsars. Our simulation 
predicts that with the present sensitivity, the $Fermi$ can detect 
 $N_{rtot}=40-61$ of the radio-selected  $\gamma$-ray pulsars 
and $N_{gtot}=36-75$
 of the $\gamma$-ray-selected $\gamma$-ray pulsars, which are more than 
the present $Fermi$ pulsars, that is, 21~radio-selected and 24~$\gamma$-ray 
selected pulsars (Abdo et al. 2010a; Saz~Parkinson et al. 2010). 
 To explain 
the difference between the observation and the model, the several reasons are 
expected; because most of the canonical $\gamma$-ray pulsars are 
located in the Galactic disk (c.f. Figure~\ref{galb}), 1) the simulated 
sensitivity of the $Fermi$ is too simple, namely, the sensitivity becomes 
much worse at  the Galactic plane ($|b|\sim 0$), 
2) a strong $\gamma$-ray background prevents the detection of the 
pulsed emissions from some pulsars, and (3) the $\gamma$-ray emissions from 
the pulsars may be missed by  source confusion with 
 the complex regions (e.g. Cygnus and Carina that lie on tangents to
 spiral arms), and/or with  the  unresolved sources that are not modeled in the
 diffuse backgrounds.  
 
As Table~2  shows, our simulation also  predicts
 that if the $Fermi$ sensitivity increase 
 factor of two from the present one, $Fermi$ can in principle detect
 $N_{rtot~\times 2}=58-89$ radio-selected  and $N_{gtot~\times 2}=79-152$ 
$\gamma$-ray-selected $\gamma$-ray pulsars.

\subsection{Implication for  future  observations}
\label{allpopu}

Finally, we discuss model implication on future observations. 
 In the observation, 
 sensitivity of the detection of $\gamma$-ray emitting pulsars really depend on 
the observed direction (that is, high or  Galactic latitude) and 
 $\gamma$-ray-selected or radio-selected pulsars, implying  the present 
 observed distribution of the pulsar characteristics   
may not represent the intrinsic population of 
the $\gamma$-ray pulsars that irradiate  the Earth.

\begin{figure}
\epsscale{.80}
\plotone{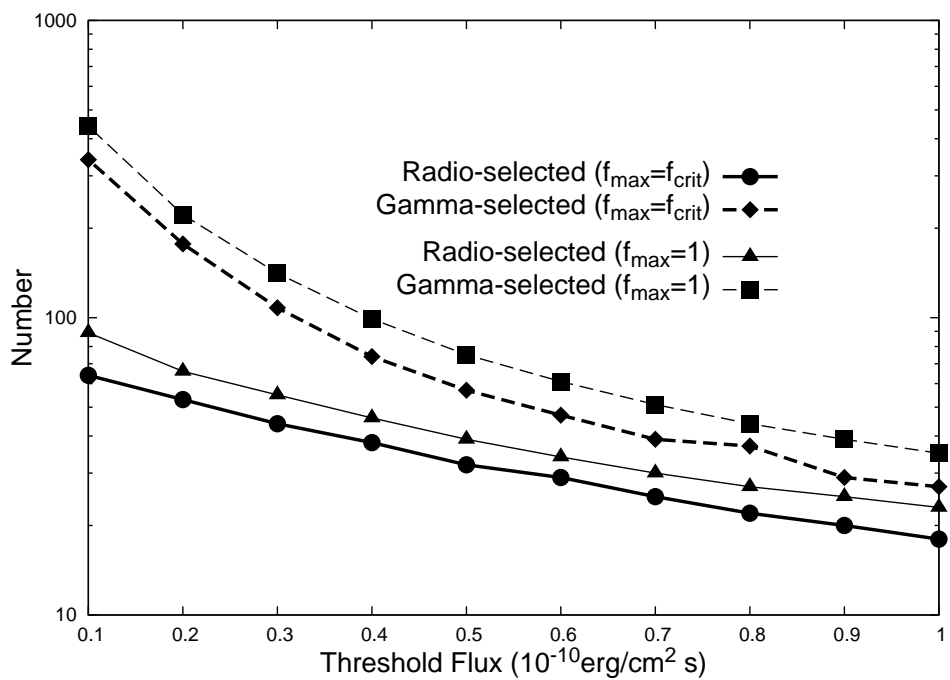}
\caption{The number of radio-selected (solid) and $\gamma$-ray-selected 
(dashed) pulsars detected within simulation as a function of the threshold 
energy flux. The thick and thin lines are results for the maximum fractional 
gap thickness of $f_{max}=f_{crit}$ and $f_{max}=1$, respectively.}
\label{number}
\end{figure}

Figure~\ref{number} summarizes number of the  
 pulsars detected  within the simulation  as a function of 
 the $\gamma$-ray  threshold energy flux. Here, because  we  do not take 
into account the dependency of the sensitivity 
 on the observed direction and on the radio-selected or $\gamma$-ray-selected
 pulsars, 
 Figure~\ref{number} presents the intrinsic number of the pulsars irradiating 
the Earth with a flux larger than the threshold flux.  
The solid and dashed lines represent
 numbers for the radio-selected and $\gamma$-ray-selected $\gamma$-ray pulsars, 
respectively.  In addition, the thin and thick lines are results for 
the fractional gap thicknesses of $f_{max}=f_{crit}$ and $f_{max}=1$, 
respectively. In the simulation, for example, 18-23 radio-selected 
(solid lines) 
and 27-35 $\gamma$-ray-selected (dashed line) $\gamma$-ray pulsars irradiate the Earth 
 with a  flux large than 
$F_{\gamma}=10^{-10}~\mathrm{erg/cm^2 s}$ measured on the Earth,
 while 64-89 radio-selected 
and 340-441 $\gamma$-ray-selected $\gamma$-ray pulsars irradiate the  Earth with  
 a flux larger than $F_{\gamma}=10^{-11}~\mathrm{erg/cm^2 s}$. 
Our simulation therefore implies that although the  minimum energy flux of 
the known $\gamma$-ray pulsars is 
approximately $F_{\gamma}\sim 10^{-11}~\mathrm{erg/cm^2 s}$,  the $Fermi$ 
has  missed  most of the $\gamma$-ray pulsar with a $\gamma$-ray flux 
larger than $F_{\gamma}=10^{-11}~\mathrm{erg/cm^2 s}$. 

\begin{figure}[h]
\epsscale{.80}
\plotone{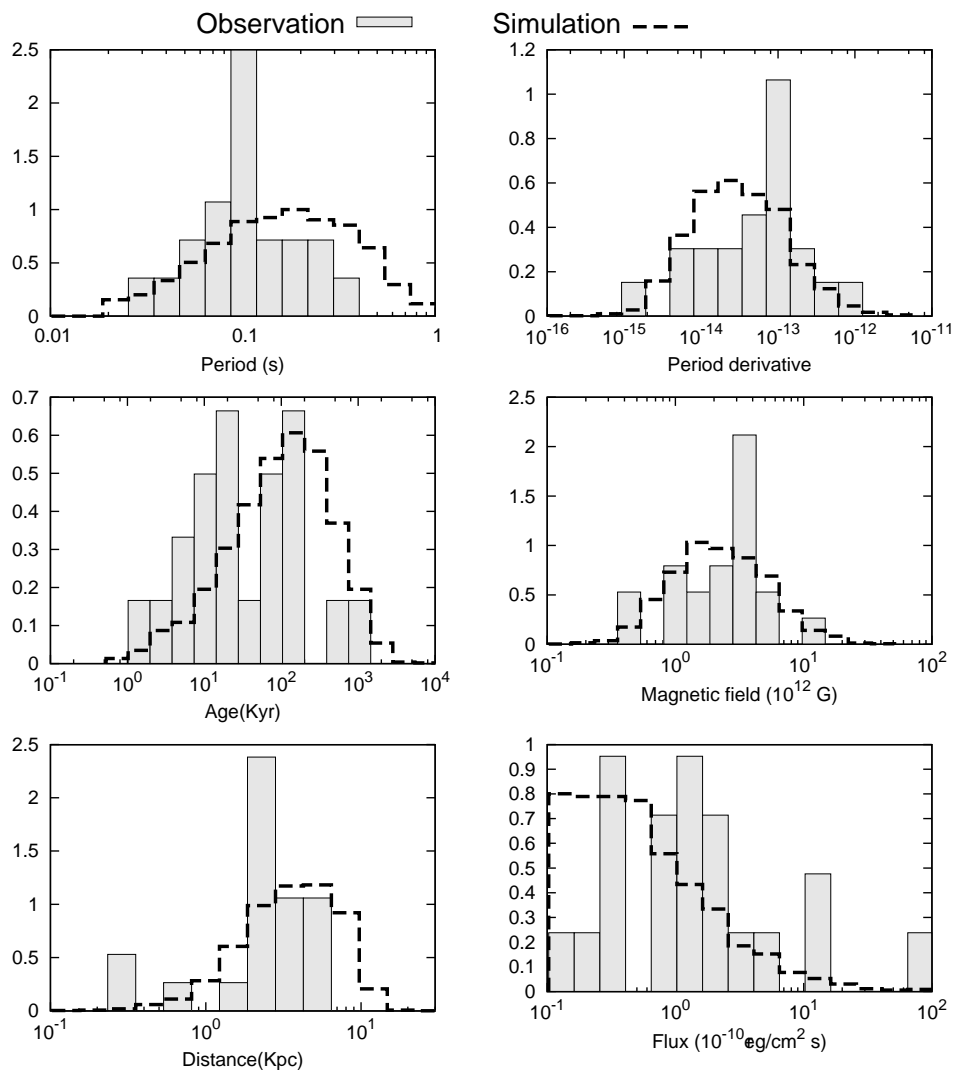}
\caption{The simulated distribution (unshaded histograms) 
of the characteristics for the radio-selected  $\gamma$-ray pulsars 
that irradiate the Earth with a flux   
$F_{\gamma}\ge 10^{-11}~\mathrm{erg/cm^2 s}$. The shaded 
histograms  present results for the $Fermi$ observation.  }
\label{comdis-rs}
\end{figure}

\begin{figure}[t]
\epsscale{.80}
\plotone{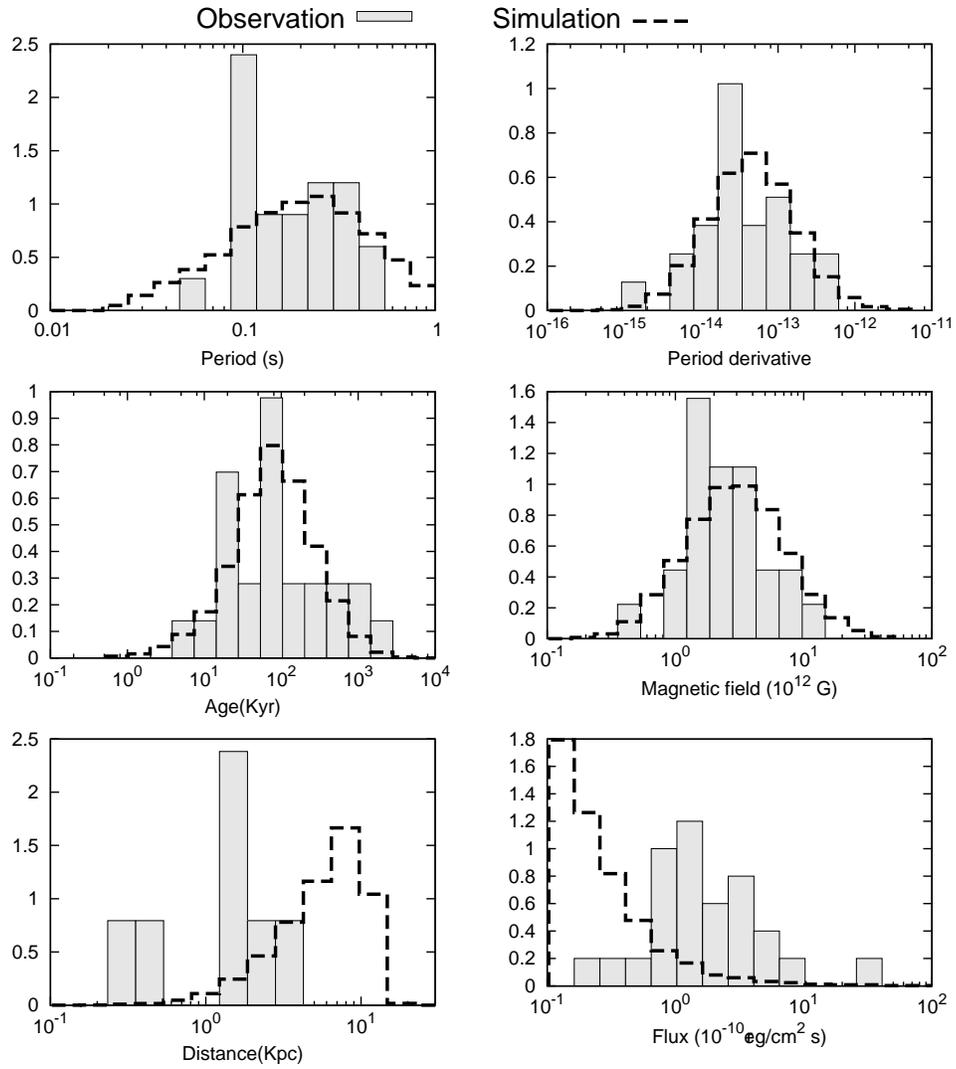}
\caption{The same with Fig.~\ref{comdis-rs}, but results for 
the $\gamma$-ray-selected  pulsars. }
\label{comdis-gs}
\end{figure}

Figures~\ref{comdis-rs} and~\ref{comdis-gs} show the distributions (unshaded 
histograms) of characteristics of the simulated radio-selected and 
$\gamma$-ray-selected $\gamma$-ray pulsars, respectively,  with a $\gamma$-ray 
flux larger than $F_{\gamma}\ge 10^{-11}~\mathrm{erg/cm^2 s}$.
In the figures we also show the observed distributions  (dashed histograms) 
by the $Fermi$. For the rotation period, the simulation predicts 
 that the future observations will detect  the radio-laud pulsars with 
longer rotation period ($P>0.5$~s) as Figure~\ref{comdis-rs} shows, 
 if  the sensitivity is improved. 
For the $\gamma$-ray-selected  pulsars, 
on the other hand, the future observations  
will  detect more  pulsars with a shorter rotation period than 0.1~s.  

We find in Figures~\ref{comdis-rs} and~\ref{comdis-gs} that 
most of the simulated pulsars distribute with a distance around 
$d\sim 10$~kpc and a flux $F_{\gamma}\sim 10^{-11}~\mathrm{erg/cm^2 s}$, while 
the distributions of the $Fermi$ pulsars have a peak at $d\sim 1$~kpc and 
$F_{\gamma}\sim 10^{-10}~\mathrm{erg/cm^2 s}$.  Therefore, our simulation 
predicts that as the  sensitivity is improved,  more and more $\gamma$-ray 
pulsars with a distance lager than $d\sim1$~kpc and a flux 
 smaller than $F_{\gamma}\sim 10^{-10}\mathrm{erg/cm^2 s}$ 
will be detected  by the future observations.  

We note that as Figures~\ref{comdis-rs} and~\ref{comdis-gs} show,
our simulation indicates  the peak of distributions for the period and
the period time derivative
come to $P\sim 0.1-0.3$~s and $\dot{P}\times 5\times 10^{-14}$~s/s, 
respectively, indicating
the $\gamma$-ray pulsars with a rotation period of $P\sim 0.1\sim 0.3$~s
 and a period time derivative of
 $\dot{P}\sim 5\times 10^{-14}~$s/s may be preferentially
 detected from  the unidentified Galactic sources.  
This information might be useful to
 narrow down the searching range of the pulse  period from
the unidentified sources to detect new $\gamma$-ray-selected  pulsars.

\section{Discussions}
\label{discussion}
\subsection{Ratio between radio-selected and $\gamma$-ray-selected pulsars}
\label{ratio}

In this section, we discuss the ratio between the radio-selected and 
$\gamma$-ray-selected $\gamma$-ray pulsars. 
With Table~3, we demonstrate how the ratio  depend on the  radio surveys. 
The first column shows the radio surveys 
used in the simulation, where M2, A2, and  P2 represent 
for Molonglo~2,  Arecibo~2 and Parks~2 surveys. 
``All'' presents   results with  all radio surveys listed in Table~1. 
In addition, we increase factor of two for the 
sensitivity of each survey.   The second and third 
column show the number of the  radio-selected ($N_{r}$) and the $\gamma$-ray 
selected pulsars that irradiate the Earth with a flux 
larger than $F_{\gamma}=10^{-10}~\mathrm{erg/cm^2 s}$ and 
$10^{-11}~\mathrm{erg/cm^2 s}$, respectively. The fourth column shows 
the results using the  $Fermi$ sensitivity of six-month observations. 
We present the  results for $f_{max}=f_{crit}$.

As Table~3 shows the  ratio between the radio-selected and radio-quiet
 $\gamma$-ray pulsars depend on the sensitivity of the radio survey.  
For example, the present simulation predicts the ratio $N_{g}/N_{r}=4$ 
using only  Arecibo~2 and Parks~2 surveys, 
 which will be consistent with the  $N_g/N_r=37/11\sim 3.4$ predicted by 
Cheng \& Zhang (1998). Because radio emissions from 
 some $\gamma$-ray selected pulsars can be detected by more sensitive
 radio survey,  the ratio using ``All'' radio survey listed in Table~1 becomes 
  $N_g/N_r\sim 1.4$, which will be consistent with the present $Fermi$ result 
$N_g/N_r\sim 17/12\sim 1.4$. 

Table~3 indicates  that the ratio depends on the threshold energy flux, 
that is, the ratio increases 
 as the threshold energy flux decreases, as second and third columns show. 
 This tendency depends on  the sensitivity of the radio survey. 
 For example, because the ``bright'' $\gamma$-ray sources are 
located relatively near  the Earth, the radio emissions from those pulsars 
can be observed by the present radio surveys if the radio emissions point 
toward the Earth. On the other hand,  dimmer  $\gamma$-ray sources
 ($F_{\gamma}\sim 10^{-11}~\mathrm{erg/cm^2 s}$) 
tend to have a more distance as indicated 
by bottom panels of Figures~\ref{comdis-rs} 
and~\ref{comdis-gs}. The flux of radio emissions from  those 
 distant $\gamma$-ray pulsars may be under the sensitivity 
of the radio surveys, and the pulsars are categorized as $\gamma$-ray-selected 
pulsars. Therefore, the ratio of the $\gamma$-ray-selected and radio-selected 
pulsars increases as the threshold energy flux decreases. 
 In fact, the present simulation predict that 
if we could detect all pulsars that irradiate the Earth with the 
radio emissions, the ratio of the $\gamma$-ray-selected 
and radio-selected pulsars  does not depend on the threshold energy flux. 

Applying the sensitivity of the $Fermi$ (fourth column), because 
the sensitivity for the radio-selected $\gamma$-ray pulsars is better 
than the $\gamma$-ray-selected pulsars, the ratio $N_g/N_r$ is in general 
smaller than those of second and third columns, whose sensitivities for 
the radio-selected and $\gamma$-ray-selected $\gamma$-ray pulsars are 
the same each other. With the all radio surveys, the ration $N_r/N_g\sim 0.9$ 
will be  consistent with the six-month $Fermi$ observations,
 $N_r/N_g\sim 24/21\sim 1.1$. 

\subsection{Comparison with unidentified $Fermi$ sources}
\begin{figure}[t]
\epsscale{.80}
\plotone{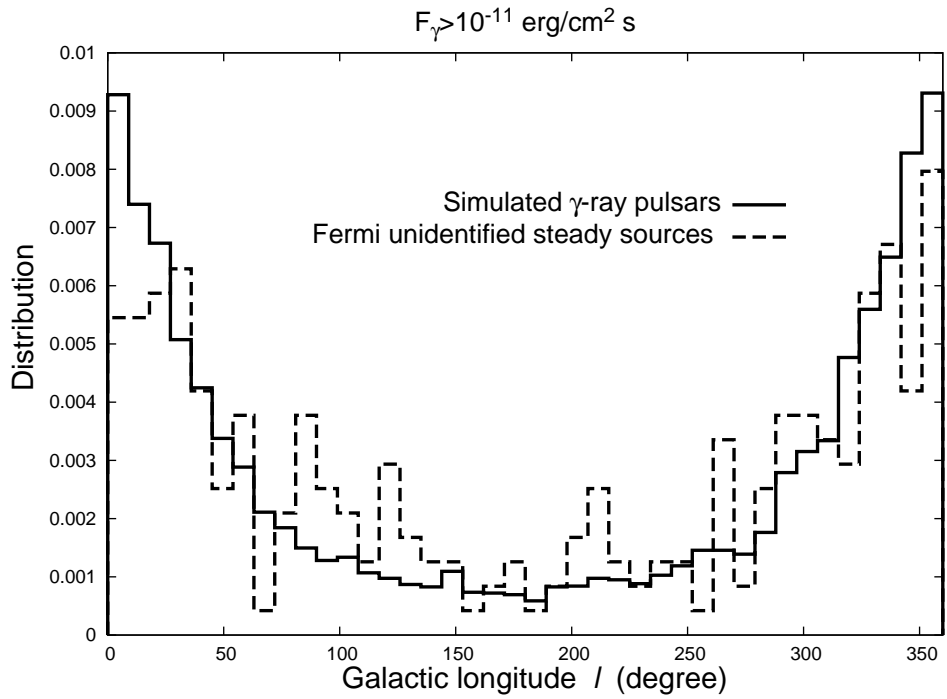}
\caption{The Galactic longitude distributions of the simulated $\gamma$-ray pulsars (solid line) and the $Fermi$ 
unidentified sources with $F_{\gamma}\ge 10^{-11}~\mathrm{erg/cm^2 s}$. 
 The data were taken from the Fermi first catalog (Abdo et al.,2010c) with 
 the criteria that the variable index less than 23.11 and the curvature 
index larger than 5.}
\label{gall}
\end{figure}

\begin{figure}[t]
\epsscale{.80}
\plotone{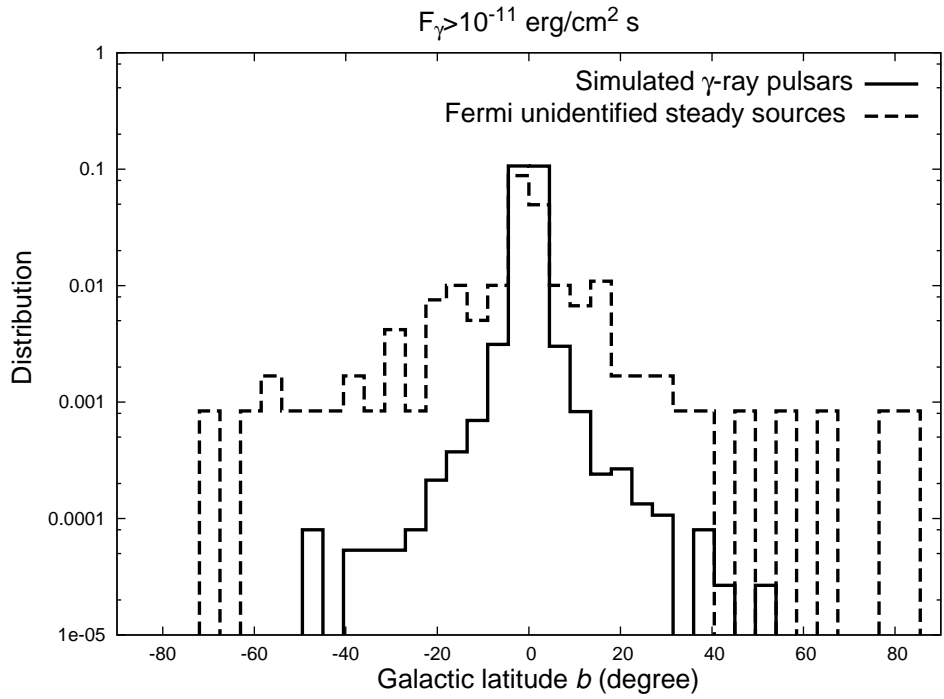}
\caption{The Galactic latitude distributions of the simulated $\gamma$-ray pulsars (solid line) and the $Fermi$ 
unidentified sources with $F_{\gamma}\ge 10^{-11}~\mathrm{erg/cm^2 s}$. The data were taken from Abdo et al. (2010c).}
\label{galb}
\end{figure}

Finally, we discuss the distributions of the $\gamma$-ray pulsars in the galaxy.
As we discussed in section~\ref{allpopu}, the present model predicts most 
of the canonical $\gamma$-ray pulsars with a 
flux smaller than $F_{\gamma}\le 10^{-10}~\mathrm{erg/cm^2 s}$ 
have been missed by
 the $Fermi$ telescope. 
 On the other hand, the $Fermi$ has found several hundreds of the unidentified 
points sources, implying  some of them 
 must be the canonical $\gamma$-ray pulsars. It will be useful to compare the 
Galactic distributions for the simulated $\gamma$-ray pulsars
 and the unidentified sources to have a hint for the origin of the 
unidentified sources. 
  Figures~\ref{gall} and~\ref{galb}, respectively, show 
  the Galactic longitude  and latitude  distributions for   the  
simulated $\gamma$-ray pulsars with 
$F_{\gamma}\ge 10^{-11}~\mathrm{erg/cm^2 s }$   (solid line) and 
 the unidentified $Fermi$ sources (dashed line).  We selected the 
unidentified sources with the criteria 
 that the observed $\gamma$-ray  flux is larger than 
$10^{-11}~\mathrm{erg/cm^2 s}$ and the variability 
 index (V) in the catalog smaller than 23.21, larger 
 than which indicates less than a 1\% chance of being
 a steady source (Abdo et al. 2010c).  In addition, we chose 
the unidentified sources with the curvature index, which is defined in 
the first catalog,  larger than 5.
 A curvature index larger than 11.34 indicates 
a less than 1~\% chance  that the power-law spectrum is a good fit for 100~MeV-100GeV. The first catalog show thats the 
 $\gamma$-ray pulsars typically have  a curvature index larger than 5.
  
 We find in Figure~\ref{gall} that the Galactic longitude distribution of the simulated $\gamma$-ray pulsars is  qualitatively consistent with that of the $Fermi$ unidentified sources, that is, both distributions have a peak  around $l=0^{\circ}$ and become  minimum around $l\sim 180^{\circ}$.   In fact, we can see that the distributions of the existing radio pulsars younger than the characteristic age of $\tau \le 5$~Myr  is also  qualitatively consistent with  that of the unidentified sources. 

In Figure~\ref{galb}, on the other hand, we can see that most of the  simulated $\gamma$-ray pulsars are located around 
the Galactic disk, that is, $|b|<5^{\circ}\sim 10^{\circ}$, 
while significant fraction of the unidentified sources are located above 
the Galactic disk.  Because we could see that  the longitudinal distribution of the simulated $\gamma$-ray pulsars is consistent with that of the existing radio pulsars younger than the characteristic age $\tau\le$5~Myr, 
our model prediction is  that most of high-latitude unidentified sources 
will not associate with  the canonical $\gamma$-ray pulsars. 
  If the majority of the unidentified sources is originated from the Galactic sources,   the millisecond pulsars may 
  be possible candidate.  Because millisecond pulsars are in general older than  the canonical $\gamma$-ray pulsars and have a 
  slower proper motion than the canonical pulsars, they are bounded by the Galactic potential. 
  It is expected that a more fraction of $\gamma$-ray emitting millisecond pulsars  is located at the higher Galactic latitude. 
  On these ground, it will be important to perform a population study of the millisecond $\gamma$-ray pulsars to reveal 
  the origin of  the $Fermi$ unidentified sources. 
  The  results of the population study for the millisecond pulsars using a Mont-Carlo  technique will be discussed in the subsequent papers. 
     
\section{Conclusion}
Inspired by  the drastically increase of the $\gamma$-ray emitting pulsars by the $Fermi$ telescope,  
we  performed a population study for the canonical $\gamma$-ray pulsars with
 a Monte-Carlo method. We applied the outer gap model with a switching of the gap closure process from the photon-photon pair-creation to the magnetic pair-creation model suggested by Takata et al. (2010). Our simulation predicts that there are 18-23 radio-selected and 26-34 $\gamma$-ray-selected $\gamma$-ray pulsars with a flux lager than $10^{-10}~\mathrm{erg/cm^2 s}$ measured on the Earth. 
Applying $Fermi$ sensitivity of six-month observations, our simulation 
predicts that $40-61$  radio-selected and 36-75 $\gamma$-ray-selected pulsars 
can be detected.  We showed that the simulated distributions 
for the various characteristics (e.g. rotation period, characteristic age) can 
successfully explain those of the $Fermi$ $\gamma$-ray 
pulsars. Our simulation also predicts that there are $\sim 64$
 radio-selected  and $\sim 340$ $\gamma$-ray-selected $\gamma$-ray 
pulsars  irradiate the Earth with a $\gamma$-ray flux larger than 
$F_{\gamma}\ge 10^{-11}~\mathrm{erg/cm^2 s}$. This indicates that 
most of the $\gamma$-ray pulsars have been missed by the present observations, although their fluxes are larger than the minimum $\gamma$-ray flux detected by the $Fermi$ so far. 
In particular, our simulation predicts that if the sensitivity of the instrument is improved, more $\gamma$-ray pulsars located with 
a distance larger  than 1~kpc and  a flux $F_{\gamma}\sim  10^{-11}~\mathrm{erg/cm^2 s}$ will be detected by in the future observations.
 The ratio of the simulated 
$\gamma$-ray-selected to radio-selected $\gamma$-ray pulsars is $\sim 1.4$ for $F_{\gamma}\ge 10^{-10}~\mathrm{erg/cm^2 s}$, while $\sim 1$  using 
6-month $Fermi$ sensitivity, although 
the ratio depends on the sensitivities of the radio surveys. 
 Finally, we compared the Galactic distributions between the simulated $\gamma$-ray pulsars and unidentified $Fermi$ sources. 
We saw that although the  Galactic longitude distribution of the simulated $\gamma$-ray pulsars is qualitatively consistent with that of the $Fermi$ unidentified sources, the simulated  latitude distribution can not explain the observations, in particular for the higher latitude sources. This may  suggest that the population study for 
the $\gamma$-ray emitting millisecond pulsars may help to understand the origin of the 
 high-latitude $Fermi$ unidentified sources.
\label{conclusion}

\acknowledgments
 We thank the useful discussions with  H.-K. Chang,
K.~Hirotani, C.Y.~Hui, B.~Rudak,
M.Ruderman and  S.Shibata. We express our appreciation to the anonymous 
referee for useful comments, which led to a much improved manuscript. 
 We also thank to Theoretical Institute
for Advanced Research in Astrophysics (TIARA) operated under Academia
Sinica Institute of Astronomy and Astrophysics, Taiwan,
which  enables author (J.T.) to use PC cluster at TIARA.
This work is supported by a GRF grant of Hong Kong SAR
Government under HKU700908P.





\clearpage




\newpage
\begin{landscape}
\begin{table}
\begin{tabular}{cccccccccccc}
\hline\hline
 & Gain &  & $T_{rec}$ & $\nu$ & $t_{i}$ & $\tau_{samp}$ & $B_{BD}$ &
 $\delta\nu$ & $l$ & $b$ &  \\
Survey & ($\mathrm{K Jy^{-1}})$ & $C_{thres}$ & (K) & (MHz) & (s) & (ms)
 & (MHz) & (MHz) & (degree) & (degree) & References \\
\hline
Molonglo 2 & 5.1 & 5.4 & 210 & 408 & 40.96 & 40 & 3.2 & 0.8 & [0,360] &
 [-85,20] & Manchester et al. (1978) \\
Green Bank 2 & 0.89 & 7.5 & 30 & 390 & 137 & 33.5 & 16 & 2 & [0,360] &
 [-18,90] & Dewey et al. (1985) \\
Green Bank 3 & 0.95 & 8 & 30 & 390 & 131 & 2.2 & 8 & 0.25 & [0,360] &
 [-18,90] & Stokes et al. (1986) \\
Arecibo 2 & 10.9 & 8 & 90 & 430 & 39.3 & 0.4 & 0.96 & 0.06 & [40,65] &
 [-10,10] & Stokes et al. (1986) \\
Arecibo 3 & 13.35 & 8.5 & 75 & 430 
& 68.2 & 0.5 & 10 & 0.078 & [35,65] & [-8,8] & Nice et al. (1993) \\
Parkes 1 & 0.24 & 8 & 45 & 1520 & 157 & 2.4 & 320 & 5 & [270,20] &
 [-4,4] & Johnston et al. (1992) \\
Parkes 2 & 0.43 & 8 & 50 & 436 & 157 & 0.6 & 32 & 0.125 & [0,360] &
 [-90,0] & Manchester et al. (1996) \\
Parks MB & 0.735 & 24 & 21 & 1374 & 2100 & 0.250 & 285 & 3 & [260,50] &
 [-5,5] & Manchester et al. (2001) \\

Jordell Bank 2 & 0.4 & 6 & 40 & 1400 & 524 & 4 & 40 & 5 & [355,105] &
 [-1,1] 
& Clifton et al. (1992) \\
Swinburne IL & 0.64 & 15 & 21 & 1374 & 265 & 0.125 & 288 & 3 & [260,50 &
 [5,15] & Edwards et al. (2001) \\

\hline
\end{tabular}
\caption{Parameters of each radio survey. }
\end{table}
\end{landscape}

\newpage

\begin{table}[h]
\begin{tabular}{cc|cc|cc|cc}
\multicolumn{2}{c}{Number} & \multicolumn{1}{c}{$N_{r,F>10^{-10}}$} & \multicolumn{1}{c|}{$N_{g,F>10^{-10}}$} & \multicolumn{1}{c}{$N_{r tot}$} & \multicolumn{1}{c|}{$N_{g tot}$} & \multicolumn{1}{c}{$N_{rtot~\times 2}$} & \multicolumn{1}{c}{$N_{gtot~\times 2} $}  \\
\hline\hline
\raisebox{-1.8ex}[0pt][0pt]{K=1} & $f_{max}=1$ & 23 & 34 & 61 & 75  & 89 & 152\\
\cline{2-8}
 & $f_{max}=f_{crit}$ & 18  & 26 & 40 & 36 & 58 & 79 \\
\hline
\end{tabular}
\caption{$N_{r,F>10^{-10}}$ and $N_{g,F>10^{-10}}$ are the number of the 
simulated  radio-selected and $\gamma$-ray-selected  pulsars with 
a flux of $F\ge 10^{-10}~\mathrm{erg/cm^2 s}$, respectively. 
 $N_{rtot}$ and $N_{gtot}$ present number of the simulated 
radio-selected and $\gamma$-ray-selected  pulsars using $Fermi$ sensitivity 
of the six-month observations; for the radio-selected pulsars, 
$F=2\times 10^{-8}~\mathrm{photons/cm^2 s}$ for the $|b|\ge 5^{\circ}$
 and $6\times 10^{-8}~\mathrm{photons/cm^2 s}$ for $|b|\le 5^{\circ}$,   
while for the $\gamma$-ray-selected pulsars, 
$F=4\times 10^{-8}~\mathrm{photons/cm^2 s}$ for $|b|\ge 5^{\circ}$ 
and $1.2\times 10^{-7}~\mathrm{photons/cm^2 s}$ for $|b|\le 5^{\circ}$.
 For $N_{rtot \times 2}$ and $N_{gtot \times 2}$, 
we increases factor of two for the $Fermi$ sensitivity.}
\end{table}

\begin{table}
\begin{tabular}{c|ccc|ccc|ccc}
\raisebox{-1.8ex}[0pt][0pt]{Survey} & $N_r$ & $N_g$ & $N_g/N_r$ & $N_r$ & $N_g$ & $N_g/N_r$ & $N_r$ & $N_g$ & $N_g/N_r$ \\
 & \multicolumn{3}{|c|}{$F_{\gamma}\ge 10^{-10}~\mathrm{erg/cm^2 s}$} & \multicolumn{3}{c|}{$F_{\gamma}\ge 10^{-11}~\mathrm{erg/cm^2 s}$} & \multicolumn{3}{c}{$Fermi$ sensitivity} \\
\hline\hline
A2, P2 & 9 & 35 & 3.9 & 24 & 372 & 15.5 & 17 & 48 & 2.8 \\
M2, A2, P2 & 13 & 28 & 2.2 & 36 & 367 & 10.2 & 26 & 40 & 1.5 \\
All & 18 & 26 & 1.4 & 64 & 340 & 5.3 & 40 & 36 & 0.9 \\
All ($\times 2$) & 20 & 26 & 1.3 & 93 & 319 & 4.2 & 47 & 34 & 0.72 \\
\end{tabular}
\caption{Ratio between  the radio-selected ($N_r$) and  
$\gamma$-ray-selected ($N_{g}$)  pulsars  
vs. radio surveys. The first column is list of 
survey, the second and the third columns present the number of the 
pulsars detected within the simulation 
with $F_{\gamma}\ge 10^{-10}~\mathrm{erg/cm^2 s}$ and $F_{\gamma}\ge
10^{-11}~\mathrm{erg/cm^2 s}$, respectively. The final 
column shows the number of the pulsars detected within the simulation using 
 the $Fermi$ sensitivity of the six-month observation.
 The results are for the beaming
 fraction of $f_{\gamma}\sim 0.4$ and the birth rate of 1.3 per century.
M2=Molonglo~2, A2=Arecibo~2 and P2=Parks~2.
 ``All'' represents all surveys listed in Table~1.
In ``All($\times 2$)'', we increase factor of two for the sensitivity of 
each survey.}
\end{table}

\newpage

\newpage
\newpage

\newpage

\newpage

\newpage 

\newpage

\newpage 

\newpage

\end{document}